\documentclass[aps,pre,twocolumn,showpacs,superscriptaddress]{revtex4-1}
\usepackage{epsf,amsmath,amssymb,verbatim,color,multirow,graphicx,times}
\begin{document}
\title{Complete trails of co-authorship network evolution}
\author{Deokjae Lee}
\affiliation{Department of Physics and Astronomy, Seoul National University, Seoul 151-747, Korea}
\author{K.-I. Goh}
\email{kgoh@korea.ac.kr}
\affiliation{Department of Physics, Korea University, Seoul 136-713, Korea}
\author{B. Kahng}
\email{bkahng@snu.ac.kr}
\affiliation{Department of Physics and Astronomy, Seoul National University, Seoul 151-747, Korea}
\affiliation{School of Physics, Korea Institute for Advanced Study, Seoul
130-722, Korea}
\author{D. Kim}
\affiliation{Department of Physics and Astronomy, Seoul National University, Seoul 151-747, Korea}
\date{\today}

\begin{abstract}
The rise and fall of a research field is the cumulative outcome of its intrinsic
scientific value and social coordination among scientists. The structure of the social component is quantifiable by the social network of researchers linked via co-authorship relations, which can be tracked through digital records. Here, we use such co-authorship data in theoretical physics and study their complete evolutionary trail since inception, with a particular emphasis on the early transient stages. We find that the co-authorship networks evolve through three common major processes in time: the nucleation of small isolated components, the formation of a tree-like giant component through
cluster aggregation, and the entanglement of the network by large-scale loops.
The giant component is constantly changing yet robust upon link degradations,
forming the network's dynamic core. The observed patterns are successfully reproducible through a new network model.
\end{abstract}
\maketitle

\section{Introduction}
At the brink of the twenty-first century, two papers heralding the
beginning of a new science of network, which were for the
small-world and the scale-free networks~\cite{ws,ba}, were
published. Since then, complex network research (CNR) has
flourished, not only as an active research field but also as a
common structural analytic framework by which the systems approach
to various complex systems in the natural, social, and information
sciences can potentially be unified
~\cite{book,caldarelli-book,internet-book,palsson,social}. Along
with the emergence of a new research field, a new research
community forms and evolves. Thus, the CNR provides a useful
example to study the spreading of a research field in terms of the
evolution of the social network behind
it~\cite{perspective,newman,toroczkai,onnela}. Specifically, we
can quantify evolutionary patterns in its initial transient
periods of the network formation, which has never been explicitly
observed. We can also portray the route through which the
co-authorship network reaches a fully-grown state.

To achieve this goal, we apply the complex network theory to study
how the CNR has developed since its inception: We first construct
the co-authorship network in which nodes are researchers
participating in the CNR and a link is made when two authors write
a paper together \cite{newman}. The weight of the link is given by the number of
papers co-authored. To track the evolution of the network, the two
aforementioned papers \cite{ws,ba} and three early review papers
\cite{rmp,advphys,siam} have been chosen. They are the
highest cited papers and regarded as pioneering papers in the CNR
field. Next, we considered all the subsequent published
papers citing any of these five papers to engage in the CNR.
According to Web of Science, there are 5,008 such papers with
information on the list of authors and publication times measured
in months, written by 6,816 non-redundant authors (in terms of
their last name and initials) for a period spanning 127 months
from June 1998 to December 2008. The co-authorship network was
constructed each month from the papers published up to that
month~\cite{exclude}. This network is a growing, weighted network.
With the data, we have performed a detailed
temporal analysis of the evolution of the large-scale structure of
the network and discuss its social implications. We particularly
emphasize the early stage of evolution, which has not been
addressed in previous studies~\cite{pref,palla}.

Our fine-scale temporal analysis in Secs.~II--IV reveals a global structural
transition of the network through three major regimes of
(I) the nucleation of small isolated components, (II) the
formation of a tree-like giant component by cluster aggregation,
and (III) the network entanglement by long-range loop formation.
The network reaches the steady state in which the mean separation
between two nodes stabilizes around a finite value.
The locality constraint, that is, new links are formed much more locally
than globally, played an important role in sustaining the network's
tree-like structure in regime (II). Here, by tree-like structure, we mean that
the network is dominated by short-range loops and devoid of
long-range connections, thus becomes a tree when coarse-grained
into the network of supernodes, corresponding in this case to
groups led by each principal investigator. 
This implies that most papers are made through in-group collaborations,
even though researchers began sharing ideas through international conferences.
If the locality effect were weak, the intermediate stage (II) would not appear.
Moreover, such a tree-like structure is a fractal 
and sustains even underneath the entangled network in late regime (II).
This structure is unveiled upon the removal of inactivated
edges, and has the same fractal dimension as in tree-like
structure. This implies that a hidden ordered structure with the same
fractal dimension underlies in the evolution process. In Sec.~V,
a model is constructed based on the empirical findings and suggests
that a structural transition in the real co-authorship network can
be understood as a percolation transition in the growth parameter space.
Finally, we will summarize the results and discuss their robustness
and implications in Sec.~VI.

\begin{figure}
\includegraphics[width=0.95\linewidth]{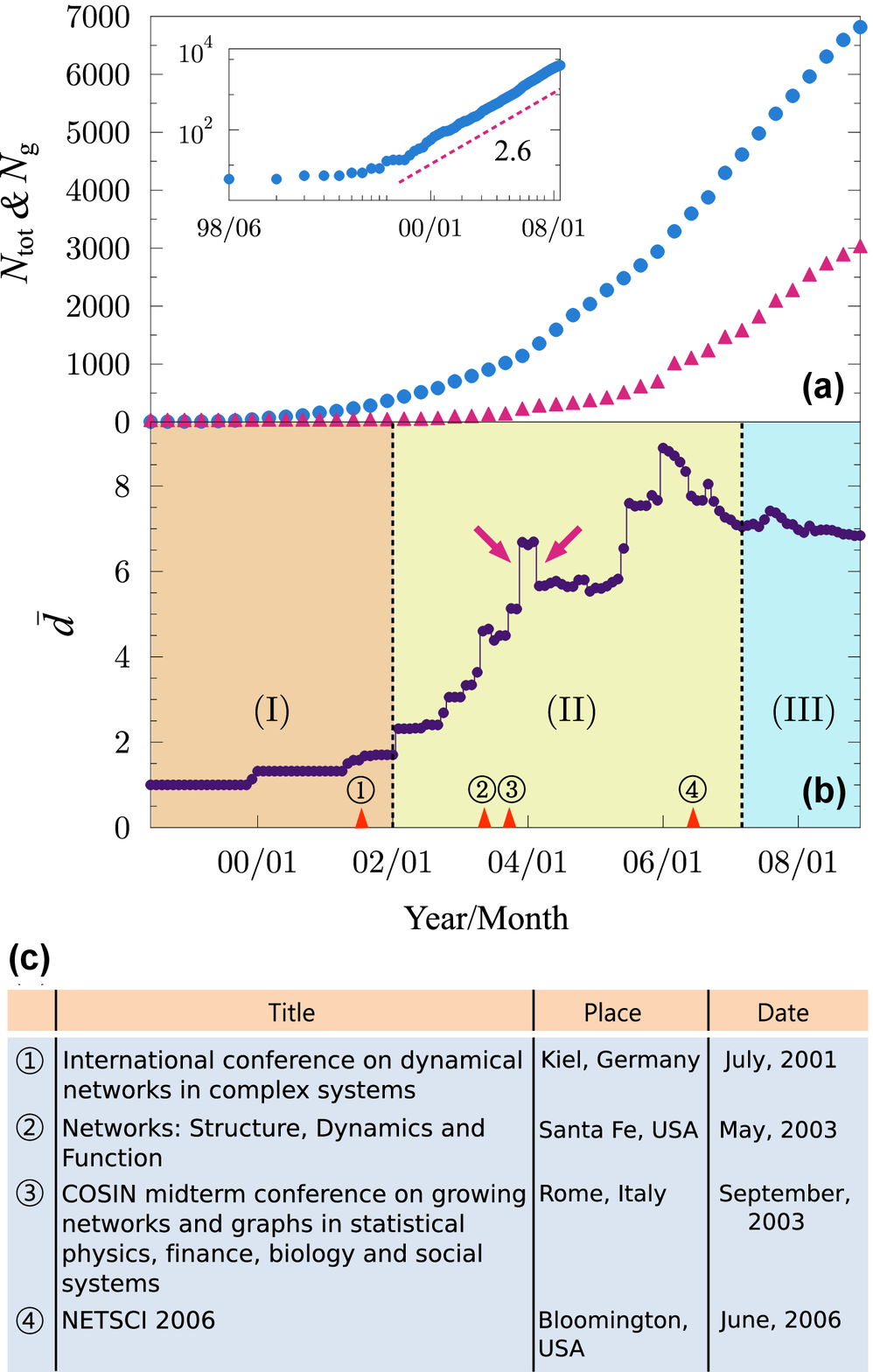}
\caption{(Color online) {\bf Large-scale evolution of the CNR co-authorship network.}
(a) The total number of nodes $N_{tot}$ (researchers;
$\circ$) and the largest component size $N_{g}$ ($\triangle$) of the network
as a function of time (quarterly). $N_{tot}$ grows in a power-law
fashion asymptotically with exponent 2.6 as shown in the inset.
(b) The mean separation $\bar d$ between a pair of nodes in the
giant component as a function of time. In the regime (I), the
giant component is small in size, and so is $\bar{d}$. It grows in
the regime (II) and exhibits intermittent jumps (indicated by the
left arrow, for example) by merging with smaller but macroscopic
components (illustrated in Fig.~2a). It also drops suddenly from
time to time (indicated by the right arrow, for example), by
making a long-range link (illustrated in Fig.~2b). In the regime
(III), the network gets entangled further and $\bar{d}$ remains
almost constant, even though the size of the giant component grows
steadily. (c) The list of the international conferences indicated
in (b).}
\end{figure}

\begin{figure}
\includegraphics[width=0.95\linewidth]{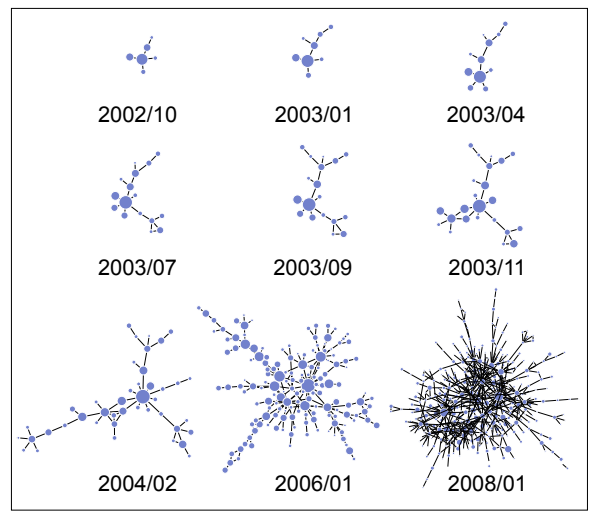}
\caption{(Color online) {\bf Structural evolution of
coarse-grained network.} The coarse-grained network obtained by
the affinity propagation algorithm \cite{affinity} is shown for
various times. A supernode in the coarse-grained network
represents a group of researchers identified by the algorithm, and
its size corresponds to the number of researchers in the group. We
note that until 2006, the coarse-grained network is effectively a
tree, devoid of long-range loops in it.}
\end{figure}

\begin{figure}[h]
\includegraphics[width=\linewidth]{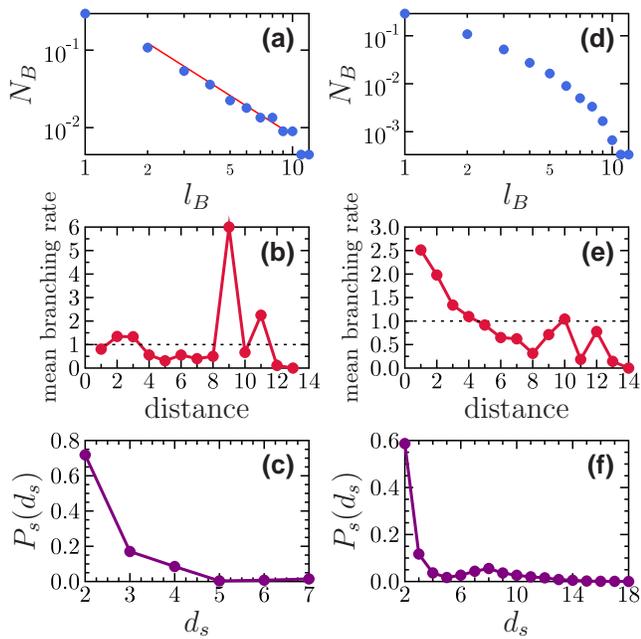}
\caption{(Color online) {\bf Fractal analysis of the giant
component.} (a) Box-covering analysis of the giant component in
February 2004. $N_B(\ell_B)$, the number of boxes needed to cover
an object with box size $\ell_B$, decays algebraically in $\ell_B$
as $N_B(\ell_B)\sim\ell_B^{-d_B}$ with $d_B\approx1.7$, indicating
that the the network is a fractal. The guideline has a slope
$-1.7$, drawn for the eye. (b) Mean branching rate as a function
of distance from a root node. To measure this, one identifies the
skeleton \cite{skeleton} of the network and its largest hub as the
root node. Then we calculate the average branching number in the
skeleton as a function of distance from the root. The plateau
around the unity value indicates its branching pattern is that of
a critical branching tree, which is a fractal. (c) The
distribution of shortcut lengths. By decomposing the network into
the skeleton and the residual links, we define the distance along
the skeleton between two nodes connected by each residual link as
the shortcut length. The shortcut length distribution of the
Feb.~2004 network has a peak at $d_s=2$ and decrease monotonically
as $d_s$ increases. In (d--f), we measure same quantities for the
network in Dec.~2008, finding that (d) the box-number decreases
exponentially with the box size, (e) the mean branching ratio does
not exhibit a plateau but decreases steadily as a function of the
distance from the root, and (f) the shortcut length distribution
becomes bimodal with an additional peak at $d_s=8$ due to the
long-range links. These results indicate that the network is no
longer a fractal in Dec.~2008. }
\end{figure}

\begin{figure}
\includegraphics[width=\linewidth]{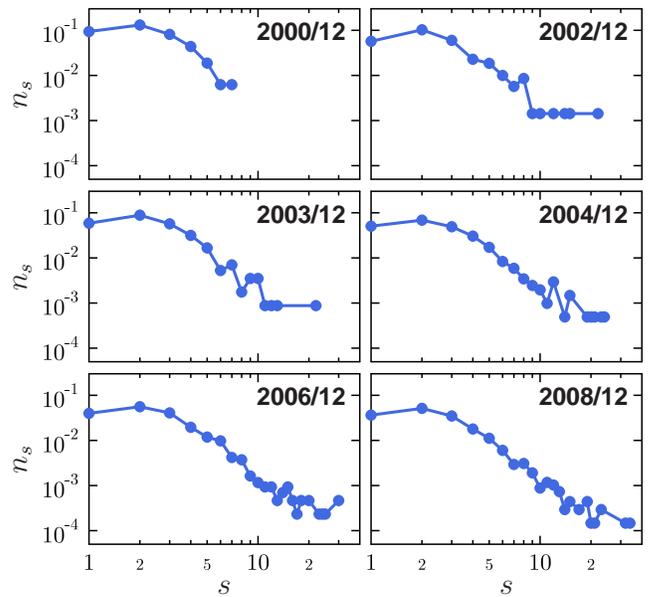}
\caption{(Color online) {\bf Component size distribution.}
Distribution of the component sizes excluding the giant component
is shown in time. We can see that the tail of the distribution
become heavier as time goes on, meaning that the heterogeneity in
component sizes increases.}
\end{figure}

\begin{figure*}
\includegraphics[width=0.7\linewidth]{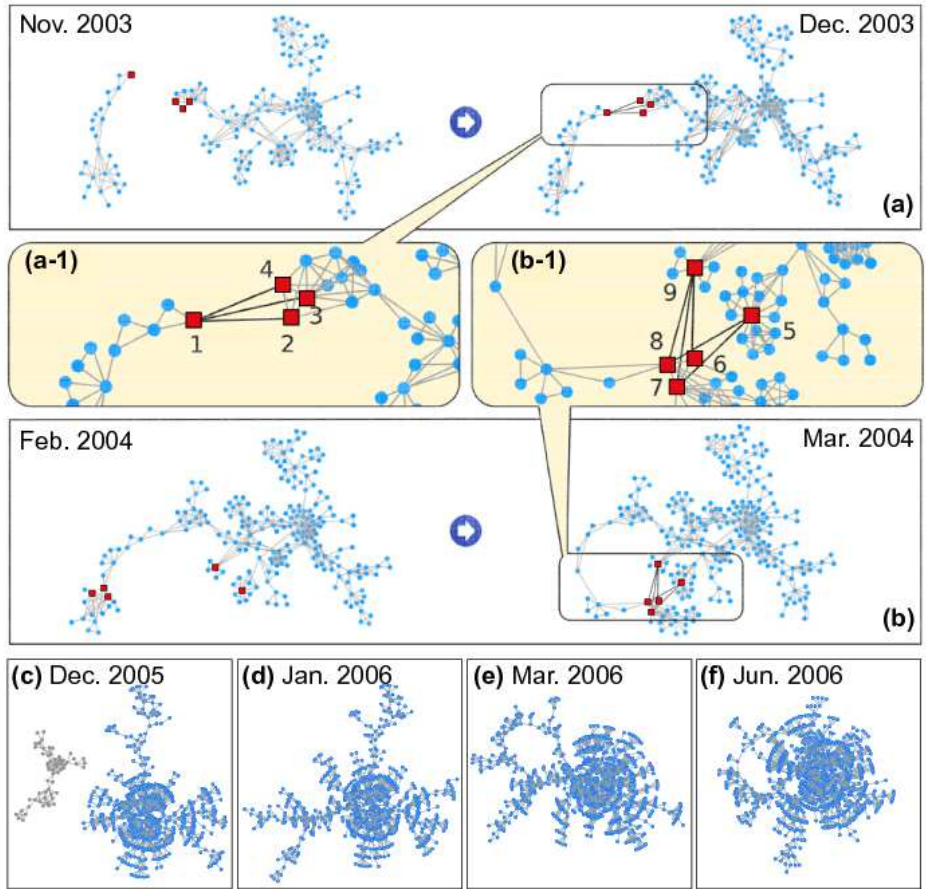}
\caption{(Color online) {\bf Large-scale changes in the giant component structure.}
(a) Between November and December 2003, the largest component
grows abruptly by merging with a smaller but macroscopic
component, leading to a big jump in the mean separation
(indicated by left arrow in Fig.~1b).
Also shown is the zoomed-in version of the merging process (a-1).
This event corresponds to the publication of the paper~\cite{paper1}
with four authors (labeled as authors 1--4 with square nodes)
from the two components.
(b) Later between February and March 2004, the largest component
acquires inter-branch link to form a long-range loop,
giving rise to a sudden drop of the mean separation (right arrow in Fig.~1b).
The zoomed-in version of the looping process is also shown (b-1).
This event is driven by the publication of two papers \cite{paper2,paper3}
with three (nodes 5, 7 and 8) and four authors (nodes 6--9), respectively,
all of whom were already in the giant component and members of COSIN project.
The links involved in the processes mentioned above are in think lines.
{(c--f) Network changes around the peak in ${\bar d}$ in January 2006.}
(c--f) A macroscopically large component has grown in Dec. 2005 (c), and it merges to
the giant component in Jan. 2006 (d), yielding a sudden increase in ${\bar d}$.
Then a large-scale loop connecting two long branches forms in Mar. 2006 (e),
and many such inter-branch links are made in Jun. 2006 (f),
yielding a sudden decrease in ${\bar d}$.}
\end{figure*}

\begin{figure}
\includegraphics[width=\linewidth]{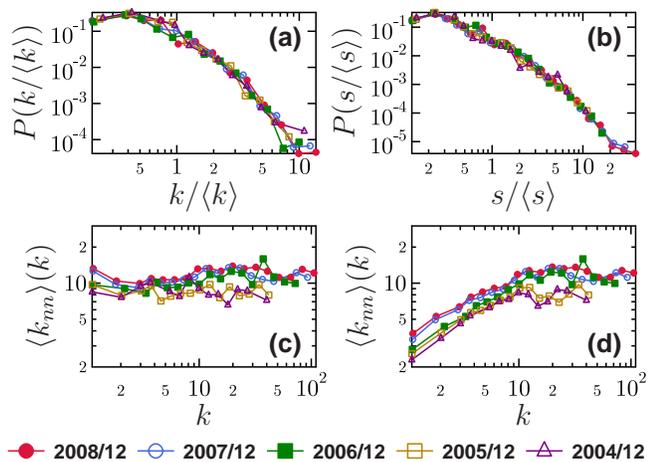}
\caption{(Color online) {\bf Basic network characteristics.}
(a--b) The degree distribution (a) and the strength distribution
(b) of the giant component. They are heavy-tailed and stationary
over time when rescaled by the average value. (c--d) The average
nearest-neighbor degree function $\langle k_{nn} \rangle$ of the
giant component (c) and of the full network (d) as a function of
degree $k$. In contrast to the full network exhibiting the
assortative mixing behavior, a typical pattern of social networks,
the giant component is almost neutrally mixed over time.}
\end{figure}

\begin{figure}
\includegraphics[width=\linewidth]{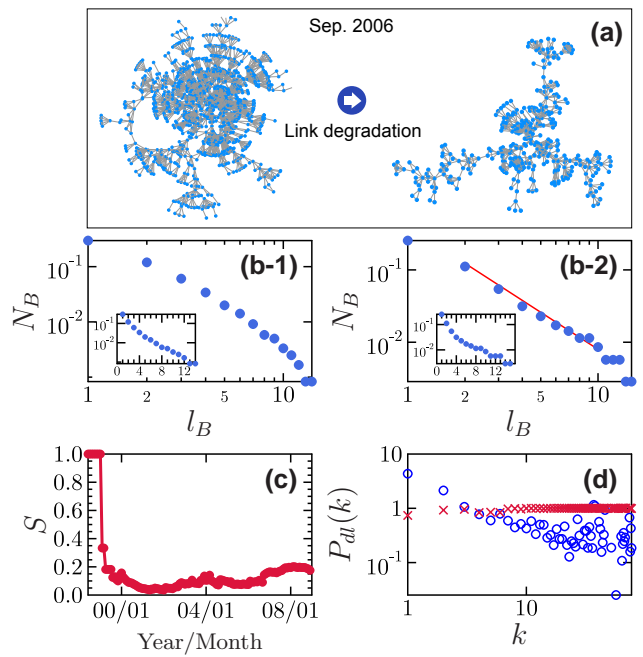}
\caption{(Color online) {\bf Link degradations.} (a) Snapshots of
the giant component without (left) and with (right) link
degradations in September 2006. The network with link degradation
(right) lacks long-range loops and thus is more tree-like, while
that without the link degradation (left) is much more entangled by
long-range loops. (b) Box-covering analysis of Sep.~2006 network
both in double-logarithmic (main) and in semi-logarithmic scale
(insets). Without link degradation $N_B$ decreases exponentially
with $\ell_B$ (b-1), thus it is not a fractal. On the other hand,
$N_B$ decays algebraically in $\ell_B$ as
$N_B(\ell_B)\sim\ell_B^{-d_B}$ with $d_B\approx 1.7$ for the LDGC
(b-2), implying its fractal structure. The guideline in (b-2) has
a slope $-1.7$, drawn for the eye. (c) The relative size $S$ of
the LDGC of the network over time. $S$ becomes stable around
$0.1\sim0.2$ after 2001. (d) The relative probability of link
deletion $P_{dl}(k)$ in the CNR co-authorship network ($\circ$)
with respect to random deletions ($\times$), as a function of the
node degree $k$. The decreasing behavior in $k$ indicates the link
removal occurs via asymmetric disassembly \cite{uzzi}.}
\end{figure}

\begin{figure}
\includegraphics[width=\linewidth]{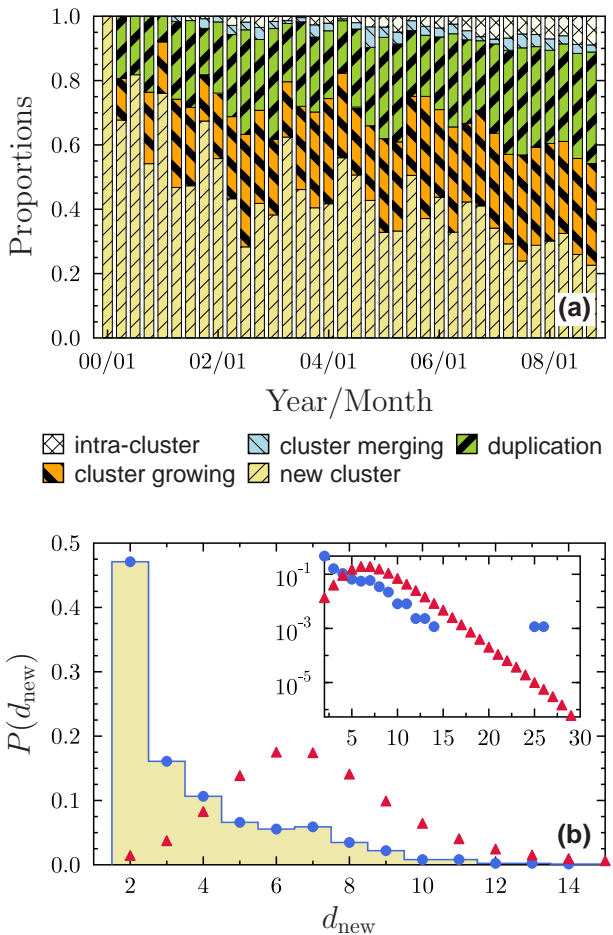}
\caption{(Color online) {\bf Microscopic link dynamics.} (a) The
proportions of new links colorcoded by the five link categories in
time (see text for details). (b) The distribution $P({d}_{\rm
new})$ of the separation of existing nodes connected by the new
links (blue circle supported by light orange bar), compared with
that for linking a random pair of nodes (red triangle). The strong
enrichment in low $d_{\rm new}$ is indicative of the importance of
locality constraint in the link dynamics. Inset: The same plot in
the semi-logarithmic scale.}
\end{figure}

\begin{figure}
\includegraphics[width=\linewidth]{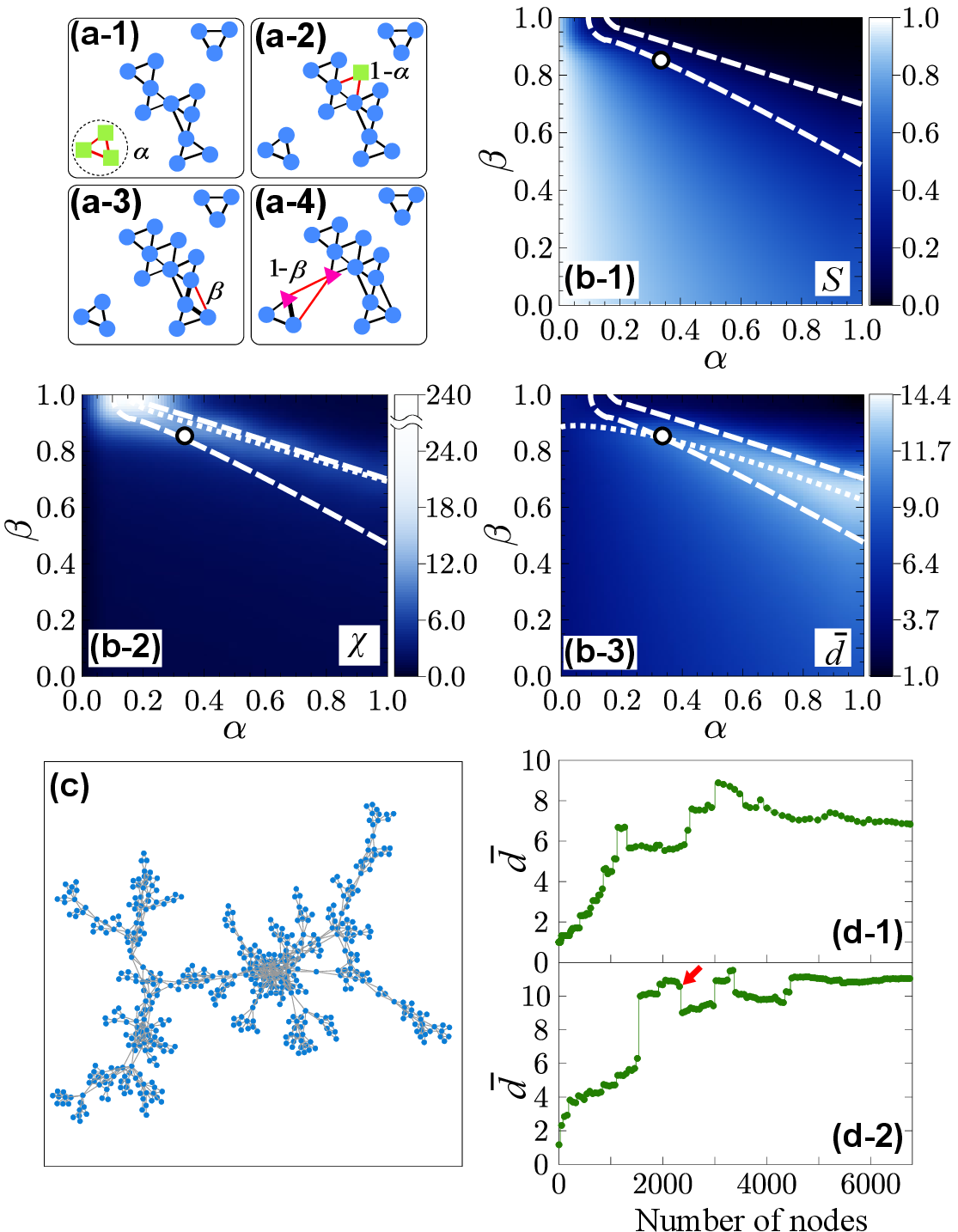}
\caption{(Color online) {\bf The network evolution model.}
(a) Schematic illustration of
the evolution rules of the network model (see text for details).
Square nodes represent the new nodes and the triangles are the
selected nodes in the rule. (b) Density plot in the
$(\alpha,\beta)$ parameter space of the relative size of the giant
component $S$ (b-1), the mean finite-component size $\chi$ (b-2),
and the mean separation of the largest component $\bar d$ (b-3)
for the network model. Numerical simulations are performed at the
0.025 intervals for both parameters and are averaged over $100$
independent runs. The approximate phase boundaries with $S=0.1$
and $S=0.3$ are drawn in dashed lines for guidance and the
empirical parameter set for the complex network co-authorship
network is indicated by a white dot. The line along the peak in
$\chi$ (b-2) and $\bar{d}$ (b-3), respectively, is drawn with the
dotted line. (c) Snapshot of the largest component of the model
network obtained with the empirical parameter set
$(\alpha=0.33,\beta=0.88)$, at the point indicated by the arrow in
(d-2) with $N_{tot}=972$ and $N_{g}=336$, exhibiting a tree-like
topology. (d) Mean separation $\bar d$ between a pair of nodes in
the giant component of the empirical network (d-1) and of a model
network with the empirical parameter set (d-2) as a function of
$N_{tot}$.}
\end{figure}

\section{Large-scale network evolution and structural properties}
Since its inception, CNR has grown steadily over the decade,
and the pace of growth has not yet started to decelerate (Fig.~1a).
The largest connected component (giant component) of the co-authorship network
has also grown in size and has reached almost a half of the total network
(Fig.~1a).
The mean separation $\bar d$ between two nodes in the giant component becomes
relatively stable to reach around $7$ after passing the intermediate regime,
during which it displays strong temporal fluctuations (Fig.~1b).
This stable behavior of $\bar d \approx 6 \sim 7$
is robust in other co-authorship networks~\cite{maldacena,randall,kpz,soc}.

The network has grown both by the expansion and merging of existing
components and by the continuous introduction of new
components. In the earliest stage (regime (I) in Fig.~1b),
small-sized components nucleate independently and their number and
size increase with time. Most of the currently highly connected
nodes (researchers) have already appeared in this regime, playing
the role of pioneers and contributing to the progress of the
field. On the brink of the intermediate stage (regime (II) in
Fig.~1b), the giant component is formed, which might be promoted
by the first international conference exclusively devoted to CNR
(\textcircled{\footnotesize 1} in Figs.~1b-c). In regime (II),
the giant component grows in a tree-like manner;
it branches out more and deeper with the passage of time, but
rarely establishes links between branches, as can be seen by
coarse-graining the network (Fig.~2) obtained by the affinity
propagation algorithm \cite{affinity}.
This can also be seen quantitatively in the distribution
of shortcut lengths, which is dominated by the peak at $d=2$ (Fig.~3c).
At this stage, the giant component is a fractal \cite{song,goh_box}
(Figs.~3a,b), with the fractal dimension $d_f\approx 1.7$ measured
by the recently introduced box-covering algorithm \cite{fractal_chaos}
and the mean branching ratio
around unity \cite{fractal_long}.
With the passage of time, such dynamics continue and
the giant component and mean separation $\bar d$ gradually increase.
Component sizes become inhomogeneous in the growth process (Fig.~4).
Such a steady growth may be promoted by
large-scale international conferences such as the one held
in Santa Fe with the purpose of bringing together scientists from
diverse disciplines to discuss network science problems across
fields (\textcircled{\footnotesize 2} in Fig.~1b-c).
However, there are a few intermittent jumps ({\it e.g.},
the left arrow in Fig.~1b),
resulting from the merging of smaller but macroscopic components
with the largest one (Fig.~5a). A large-scale loop does not appear
until 2004, and it is formed by the long-range inter-branch link
(Fig.~5b). Such a long-range loop formation can be monitored by
the sudden drop in the mean separation of the giant component
({i.e.}, the right arrow in Fig.~1b). This change
can also be monitored by examining the size of largest bi-connected component.
Furthermore, before this long-range loop formation the ratio
between the largest bi-connected component size to the largest singly-connected
component size tends to decrease, implying the tree-like growth of the
largest component during the period.
This event is a consequence of the first major multinational project
devoted to CNR in Europe (COSIN) (\textcircled{\footnotesize 3} in Figs.~1b,c).
Another prominent example is the peak in January 2006 (Figs.~5c,d).
Since then, an increasing number of large-scale loops have been formed,
resulting in an increasingly entangled and interwoven giant component structure
and the network has made the transition into regime
(III), in which the network properties such as the mean separation
become stable, despite the steady growth of the giant component.
This transition into a stable research field may be epitomized by
the establishment of a {\it regular} international conference
gathering researchers from various multidisciplinary fields
(International Workshop and Conference on Network Science
(NetSci); \textcircled{\footnotesize 4} in Figs. 1b,c).

Temporal evolution of motif contents reveals
the global structural changes from a different viewpoint.
We observed that the motifs with one triangle
begin to be significant approximately from the beginning of the intermediate
regime (II), whereas the motifs containing two triangles do so
around June 2005, at which the mean separation exhibits a drastic jump.
In this way, the temporal evolution of motif contents is related to
that of the mean separation and complements the global evolution picture.

The co-authorship network exhibits heavy-tailed behaviors in the degree
(number of links a node is connected to) and strength (the sum of the weights
of links a node has) distributions, which become robust over time
(Figs.~6a,b). The degree-degree correlation within the giant component
is almost neutral or weakly assortative, in contrast with
the assortative behavior observed for the full network and other
social networks (Figs.~6c,d) \cite{friendship,mobile}.

\section{Effect of link degradations}
Social ties decay in strength over time in the absence of reinforcement.
Co-authorship links may be no longer active if the collaboration
ceased long ago.
Thus, to ensure that the generic features remain robust, it is informative
to examine how the overall
network structure is affected in the presence of a link degradation process.
The central question would be whether the giant component persists
to support the integration of the research field.
To this end, for each month, we removed
all the links that had not been re-activated during the previous two years--a
typical postdoctoral contract period.

The link degradation process significantly affects network
configurations because many links become inactive in the end (Fig.~7a);
for example, 86\% of the links formed up to the year 2006 eventually
disappeared before the end of 2008 according to the two-year
inactivation rule. However, the giant component not only persists
upon degradation, but is also more stable, in the sense that its
relative size $S$ has been stable at $\approx$10\% of the total
network since 2000 (Fig.~7c). At the same time, the link-degraded
giant component (LDGC) is highly dynamic, in that its members
constantly change over time. At the end of 2008, 1,195 nodes
formed the LDGC, among which only 272 were the LDGC members in
December 2006 composed of 727. This indicates that the CNR is
still a vigorous field \cite{palla}. The LDGC exhibits a tree-like
structure throughout the observation period,
implying that such a tree-like spanning component
structure exists to provide a dynamic backbone underlying the
complex original interwoven network. Furthermore, the LDGC
(Fig.~7a) topologically resembles the original giant
component in regime (II) (Fig.~5a), and their
fractal dimensions are the same as $d_f\approx 1.7(1)$ (Figs.~3d and 7b).

Link degradation properties indicate the future prospects
of the research field. Based on the evolutionary trajectory observed,
the CNR has passed its initial transient growth period and has now settled into
a steady growth regime with stationary topological properties such as the
degree distribution $P_k(k)$ and the mean separation $\bar{d}$,
even taking the link degradation process into consideration.
Moreover, the link degradation process in the co-authorship network
occurs in a manner that is consistent with the so-called asymmetric
disassembly \cite{uzzi}, where the probability of link degradation decreases
with the degree of the connecting node (Fig.~7d).
Given that asymmetric disassembly provides structural
robustness in a declining network \cite{uzzi}, the current steady growth
of the CNR co-authorship network backed up
by the asymmetric disassembly implicates the integration
and stability of research discipline in the future,
even after it eventually enters into the network saturation stage.

\section{Microscopic link dynamics}
To understand the microscopic mechanisms responsible for the large-scale
evolution pattern observed, we focus on the link dynamics.
We categorize each new link into five classes
depending on the nature of the nodes that it connects
and measure their relative frequency in the link dynamics (Fig.~8a).
They are
{i)} the duplicate link, connecting two nodes already linked,
{ii)} the intra-component link, connecting two unlinked nodes
in the same component,
{iii)} the cluster-growing link, connecting an existing node
in a component to a new node, thereby resulting in an
incremental growth of the component,
{iv)} the cluster-merging link, connecting two nodes
in different components, and
{v)} the new-cluster link, connecting two new nodes
to introduce a new component.

Among them, the duplicate link, the cluster-growing link, and the
new-cluster link are found to be of high frequency, each
constituting approximately a quarter to a third of all the links.
The remaining two classes, the cluster-merging
link and the intra-cluster link are far less frequent, comprising
2.8\% and 4.7\% of all links, respectively.
Although infrequent, the latter two classes are the driving forces
of major large-scale structural changes: The former provides the
punctuated growth of components by merging existing macroscopic
components, while the latter can introduce long-range loops that
entangle the connectivity structure.

We found that the existing nodes play an equally important role as
that of the new nodes: Among nodes connected by new links,
approximately half of them are existing nodes (51\%).
We also found a clear signature of the locality effect (i.e.,
the tendency of nodes in proximity to link with each other),
which manifests itself as a strong enrichment of the links connecting nodes at shorter
separations, compared to random linkages without such a locality constraint.
For example, about 47\% of the links between existing nodes are found to be
separated by two links before linking, that is, they connect ``friends of a friend'',
compared to 1.5\% for random linkages (Fig.~8b).
It is this locality-constrained link formation that is responsible for
the tree-like growth dominating the early structure of the network.

\begin{figure}
\centering
\includegraphics[width=\linewidth]{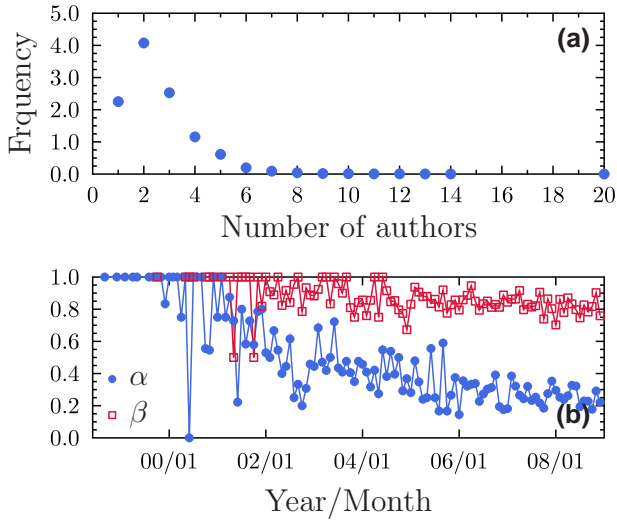}
\caption{(Color online) {\bf Empirical model parameters.} (a)
Distribution of the number of authors per paper calculated from
all 5,008 papers is plotted. The average value is obtained to be
$2.9$. Note that we excluded the papers with more than 20 authors.
(b) The network model parameters $\alpha$ and $\beta$ are measured
from the CNR publication data. The parameter $\alpha$ changes
slowly in time, the global average being $\approx 0.33$, while the
parameter $\beta$ is relatively stable as $\approx0.85$.}
\end{figure}

\begin{figure}[t]
\includegraphics[width=0.90\linewidth]{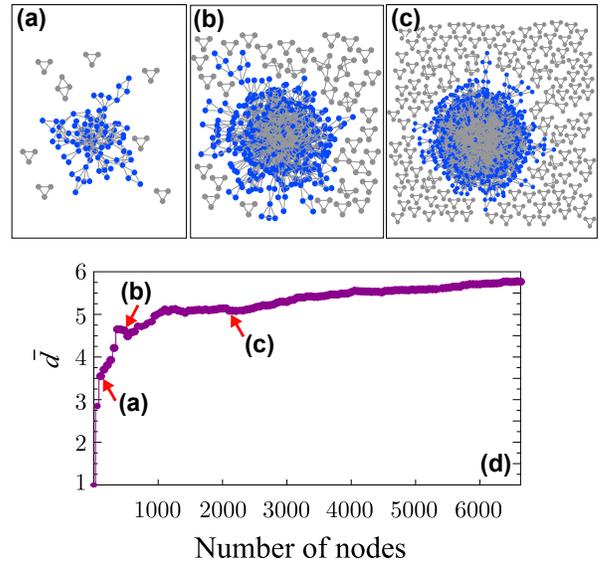}
\caption{(Color online) {\bf Evolution of the model network without locality constraint.}
%with $\beta$ much lower than the empirical value with $\alpha=0.33$ and $\beta=0.2$.}
The network snapshots (a--c) and the mean separation $\bar{d}$ (d)
over time obtained from a typical run of the network model with
parameters $\alpha=0.33$ and $\beta=0.2$. Very weak locality
constraint is imposed to the model network with this low value of
$\beta$, and it is unable to maintain the initial tree-like growth
stage but gets interwoven immediately. Therefore one cannot
observe the macroscopic-scale cluster aggregation process except
the incremental growth of the giant component.}
\end{figure}

\section{Network evolution model}
We model the co-authorship network evolution by incorporating
the observed microscopic link dynamics.
The network model is built upon a number of previous
network models \cite{callaway,dm,holme,guimera,kertesz},
with relevant growth ingredients such as the preferential attachment-based
internal link formations \cite{dm}, the triad formation due to
locality constraint \cite{holme}, and the team-based evolution \cite{guimera}.
Although all these ingredients are found relevant in the co-authorship
network evolution, none of them alone can account for the whole process.
Therefore, we combined ingredients from these previous models and incorporate
them into the combined model with additional parameters for the relative frequencies of these processes.
In this combined network evolution model which is schematically depicted in Fig.~9a,
the model network evolves by a node dynamics and a link dynamics, according to the following rules
applied at each time step: {i)} With the probability $\alpha$,
a new component with three connected nodes (a triangle) is introduced, representing a new group.
The average number of authors per paper is observed to be approximately three
(Fig.~10a). {ii)}
With the complementary probability $1-\alpha$, a new node is added, and then it selects an existing component
in proportion to the component size and connects to a node in the component
chosen in proportion to the degree (preferential attachment \cite{ba}),
as well as to a randomly chosen neighbor of that node, forming a triangle.
{iii)} Independently, with the probability $\beta$,
a randomly chosen node links to a random neighbor of its neighbor (a friends' friend).
{iv)} With the complementary probability $1-\beta$,
a triangle is formed by a random pair of nodes with separation
larger than two and one of their nearest neighbors.
Here, the parameters $\alpha$ and $\beta$ control the influx of new components
and the strength of the locality effect, respectively.

\subsection{Model simulation results}
We run the network model up to the size $N=6800$ for various
growth parameters $\alpha$ and $\beta$ and calculate the general characteristics
of the model network, specifically the fraction of nodes in the giant component $S$ and
the average size of the finite (non-giant) components $\chi$ used in the percolation study.
Here, $S\equiv 1-\sum_s sn_s$ and $\chi \equiv \sum_s s^2 n_s/\sum_s sn_s$,
where $n_s$ is the fraction of $s$-sized component and the summation runs
over finite components. In addition, we measure the mean
separation $\bar{d}$ between two nodes in the giant component, which may be
analogous to the correlation length used in the percolation theory.
$S$ decreases monotonically in both $\alpha$ and $\beta$ (Fig.~9b-1),
quite abruptly in the region bounded with dashed lines.
$\chi$ (Fig.~9b-2) exhibits a peak behavior along the same region,
denoted by dotted line. Both behaviors suggest a percolation transition-type
event occurring across the region in the parameter space.
The mean separation $\bar{d}$ (Fig.~9b-3) also displays a peak behavior
along the same region in a large $\alpha$ regime, denoted by dotted line,
establishing that the giant component is tree-like in the percolation transition region.

Having understood the generic behavior of the network model,
where does the real co-authorship network reside in the parameter space?
We measure the parameter set from the empirical data, finding that while
$\beta$ is steady at $\beta_{\rm em}\approx 0.85$, $\alpha_{\rm em}$ depends
on time (Fig.~10b). We estimate it roughly by taking the average over time to be
$\alpha_{\rm em}\approx 0.33$. Interestingly, the measured parameter set (indicated
by a white dot in Figs.~9b) is located within the percolation transition region.
This implies that the network achieves a balance between the continuous influx
of new isolated components and the formation of global connectivity.
The tree-like giant component may be rooted in the fact that
most research groups tend to work independently and rarely collaborate with other group
members; however, the hub group members perform out-of-group collaborations
more actively by locating themselves near the center of the network.

The configurations generated from the network model with the empirically measured
parameter set successfully reproduce the observed general large-scale structural
features, such as the tree-like growth of the giant component (Fig.~9c)
as well as a strong fluctuation in the mean separation $\bar{d}$
in the early time regime, followed by its stabilization through the network
entanglement by long-range loops in the later time regime (Fig.~9d).
Outside of the empirical parameter point, the model network evolves in different ways.
Notably, when the locality effect is weak (small $\beta$),
the network is unable to maintain the initial tree-like growth
stage and quickly forms a hairball-like interwoven
structure, without the macroscopic-scale cluster aggregation process, as in most
of the unconstrained random growth models (Fig.~11).
Furthermore, generalization of the model with more complex
component structure such as a mixture of dimer and trimer in the growth
rule does not affect the main results.
Thus, the current simple network model appears to address
the essential mechanisms underlying the evolution
of real networks in a minimal way.

\begin{figure}
\centering
\includegraphics[width=\linewidth]{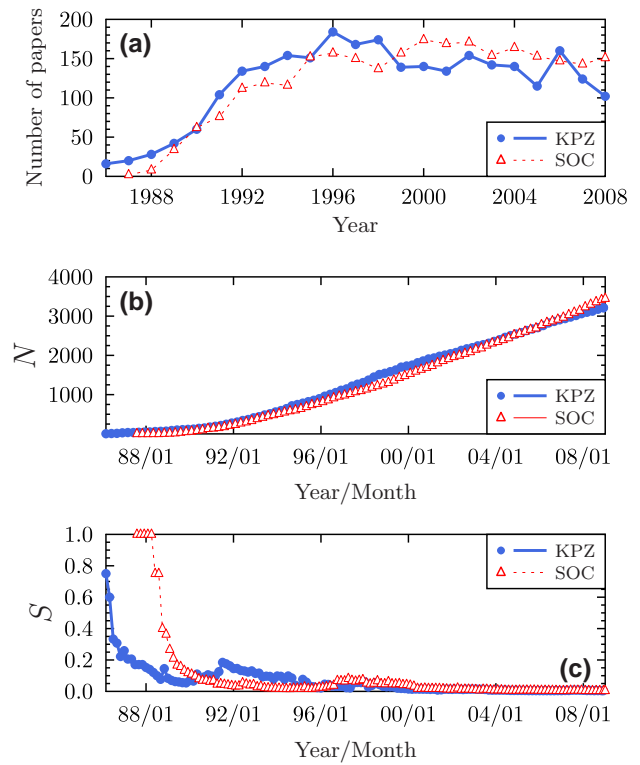}
\caption{(Color online) {\bf Co-authorship network evolution of
declining subjects.} (a) Plot of the numbers of papers published
each year versus time for two fields, fractal surface
growth obtained with Ref.~\cite{kpz} and
self-organized criticality with Ref.~\cite{soc}
as the seed papers. They show steady-state behaviors
after around the year 1996. (b) Plot of the numbers
of distinct authors contributed to the papers in (a)
versus time for the two subjects, which show monotonically
increasing behaviors. New authors steadily join the
coauthorship networks. (c) Plot of the relative sizes
$S$ of the LDGC of the co-authorship networks
in both fields Compared with the CNR network,
$S$ of these two fields degenerate below $0.01$, an order of
maginitude smaller, meaning that the giant component is
practically no longer present.}
\end{figure}

\section{Discussion}
In the past 10 years, more than 5,000 papers have been
published on the subject of the CNR by approximately 6,800
researchers. Using the Web of Science database, we have traced
those papers for 127 months.
The evolution exhibits the
percolation transition through which the giant component forms to
establish global connectivity under the continuous supply of
non-inbred new members into a society. The co-authorship network
currently appears to maintain diversity without sacrificing the
internal evolution within groups, by a moderate value of model
parameter $\alpha_{\rm em}\approx 0.33$. However, in order to form
a stable research discipline, both the growth and the
connectedness are important. In this respect, the existence of the
giant component spanning the system, even with link degradation,
would represent the maturity of the subject in that it allows the
exchange of ideas through the body of the community by occasional
unconstrained collaborations that overcome the prevailing locality
effect. Such formation of a giant component would correspond to
emergence of the so-called invisible college~\cite{guimera}.
However, such a dominant college supported by the core tree-like
giant component is shown to be highly dynamic, raising the
possibility of continuous diversity
and variations in the leading ideas and research trends.
What is remarkable is that these evolutionary patterns appear
robustly in other systems. In addition to the CNR, we chose two
new topics in theoretical high-energy physics, the
Anti-DeSitter/Conformal Field Theory duality conjecture~\cite{maldacena}
and the Randall-Sundrum model~\cite{randall}, and confirmed
that their evolutionary pattern is similar to what we discussed for the CNR.
On the other hand, when we consider a research field that is fading
away, we get different patterns. For example,
the core giant component spanning the co-authorship network disappears
after link degradation, as observed in the cases
of the fractal surface growth and
the self-organization criticality, triggered by the papers~\cite{kpz}
and \cite{soc}, respectively (Fig.~12). These two subjects have passed
their heydays in 1990s, and now their LDGC has degenerated even though
new papers are published in a steady rate and 
the new comers still continue to enter the fields.
Thus the relative size of the co-authorship LDGC may be an indicator
of the current state of a research field.
Even though there may be other subject-specific factors that
are responsible for some of the observed properties and therefore
comparison between different fields has to be interpreted with care,
our finding that they exhibit many shared patterns suggests
the validity of common evolutionary mechanisms explored in this work.

We have shown that two microscopic mechanisms, the continuous
influx of new nodes and groups and link formations strongly
constrained by the locality effect, underly the observed
co-authorship evolution pattern. Moreover, the current state of
the network was found to be nearly critical from the perspective
of percolation theory. An important remaining question is how the
system has located itself in such a delicate state. One appealing
answer might be that it has done so in a self-organized way,
calling for further studies in this direction.

\section*{ACKNOWLEDGEMENTS}
This work is supported by Mid-career Researcher Program through NRF grants
funded by the MEST (No.~2010-0015066) (to BK) and (No.~2009-0080801) (to K-IG),
and by KRCF (to DK).


\begin{thebibliography}{99}
\bibitem{exclude} We excluded 11 papers with more than 20 authors, most of which were from large-scale biology experiments such as the human genome sequencing project.
\end{thebibliography}
\end{document}